\documentclass[%
 reprint,
 amsmath,amssymb,
 aps,
]{revtex4-1}

\usepackage{graphicx}
\usepackage{dcolumn}
\usepackage{bm}
\usepackage{hyperref}
\usepackage{subfigure}  
\usepackage{xcolor}  
\usepackage[normalem]{ulem} 
\usepackage{amsmath,amssymb,mathrsfs,dsfont}

\makeatletter

\newcommand{\Rmnum}[1]{\expandafter\@slowromancap\romannumeral #1@}
\makeatother

\begin{document}

\preprint{APS/123-QED}

\title{Contagion processes on the static and activity driven coupling networks}

\author  {Yanjun Lei $^{1,2}$}
\author {Xin Jiang$^{2,3}$}
\email{jiangxin@buaa.edu.cn}
\author{Quantong Guo $^{2}$}
\author{Yifang Ma$^{1,4}$}
\author{Meng Li$^{2}$}
\author{Zhiming Zheng$^{1,2}$}%
\affiliation{%
$^{1}$School of Mathematical Sciences, Peking University, Beijing, 100871, China\\
$^{2}$LMIB \& School of Mathematics and Systems Science, Beihang University, Beijing 100191, China\\
$^{3}$Department of Engineering Sciences and Applied Mathematics, Northwestern University, Evanston, IL 60208, USA\\
$^{4}$Center for Complex Network Research, Department of Physics, Northeastern University, Boston, MA 02115, USA}

\date{\today}

\begin{abstract}
The evolution of network structure and the spreading of epidemic are common coexistent dynamical processes. In most cases, network structure is treated either static or time-varying, supposing the whole network is observed in a same time window. In this paper, we consider the epidemic spreading on a network consisting of both static and time-varying structures. At meanwhile, the time-varying part and the epidemic spreading are supposed to be of the same time scale. We introduce a static and activity driven coupling (SADC) network model to characterize the coupling between static (strong) structure and dynamic (weak) structure. Epidemic thresholds of SIS and SIR model are studied on SADC both analytically and numerically with various coupling strategies, where the strong structure is of homogeneous or heterogeneous degree distribution. Theoretical thresholds obtained from SADC model can both recover and generalize the classical results in static and time-varying networks. It is demonstrated that weak structures can make the epidemics break out much more easily in homogeneous coupling but harder in heterogeneous coupling when keeping same average degree in SADC networks. Furthermore, we show there exists a threshold ratio of the weak structure to have substantive effects on the breakout of the epidemics. This promotes our understanding of why epidemics can still break out in some social networks even we restrict the flow of the population.
\end{abstract}

\pacs{Valid PACS appear here}
\maketitle

\section{\label{sec1}Introduction}

Networks modeling plays a key role in identifying structural properties and analyzing the epidemic spreading on networks\cite{Imm1,Imm2,Imm3,Imm4,Imm5,Imm6,Imm7}. Meantime, it has been acknowledged for decades that connectivity patterns is an important factor in determining the properties of dynamic process\cite{Imm8,Imm9,Imm10,Imm11,Imm12}. Since the structure in real networks usually varies during the contagion spreading, more and more attentions have been paid on co-existence dynamic pictures where the time scale of network evolution $\tau_{G}$ differs from the time scale of dynamic process $\tau_{DP}$ \cite{Imm13,Imm14,Imm15,Imm16,Imm17,Imm18,Imm19,Imm20,Imm21,Imm22,Imm23,Imm24,Imm25}.

Generally, the difference between $\tau_{G}$ and $\tau_{DP}$ usually results in different kinds of spreading progresses. Most results obtained over recent years are mainly focused on two limit cases. In one case, $\tau_{G}\gg\tau_{DP}$, which means that the evolution of networks is much slower than that of dynamic processes. These networks are considered as of static structures. In the other case, $\tau_{G}\ll\tau_{DP}$, which means the structure changes much faster. Networks in this case are modeled as annealed ones. In fact, models based on these two limits are appropriate in analyzing static networks or annealed networks correspondingly, such as technological networks\cite{Imm10}, transmitted diseases networks \cite{Imm11}. However, these models are not suitable for time-varying networks where $\tau_{G}\sim\tau_{DP}$ \cite{Imm16,Imm17,Imm22,Imm23,Imm24}. The reason is that when $\tau_{G}\sim\tau_{DP}$, there actually exists uncontrolled biases in the characterizations of the contagion process \cite{Imm13,Imm14,Imm15,Imm16,Imm26,Imm27,Imm28,Imm29,Imm30,Imm31,Imm32,Imm33,Imm34,Imm35,Imm36,Imm37,Imm38}.
In summary, for most results, the epidemic spreading processes are usually studied separately on single structured networks, either on static \cite{Imm39,Imm40} or time-varying networks\cite{Imm13,Imm14,Imm15,Imm16,Imm17}.

In fact, for temporal networks, such as social networks, the structures usually show the so called chimera phenomenon. On one hand, some links are dynamical and temporal during network evolution and can be treated as weak connections between nodes (We call these links as weak structure in the following discussion, $\tau_{G}\sim\tau_{DP}$). On the other hand, some links are static and invariant which we treat as strong structures ($\tau_{G}\gg\tau_{DP}$). For example, when analyzing disease spreading on social networks, we find the social links can be generally divided into two kinds of groups: strong connections between family members or close friends and weak connections between strangers. Links connecting family members or close friends keep static, while links between strangers are time-varying since one may encounter various strangers everyday. We notice that some researches also study time-varying networks where links have memories\cite{Imm21}. In fact, in this case, we can also consider the links with memory to be strong and the memoryless links to be weak. These two kinds of links constitute strong structure and weak structure, respectively.

In this paper, we propose a static and activity driven coupling (SADC) model to study epidemic spreading processes on networks consist of both strong structures($\tau_{G}\gg\tau_{DP}$) and weak structures($\tau_{G}\sim\tau_{DP}$). We use the activity driven model\cite{Imm13} to describe the weak structure, where each node $i$ is assigned with a certain activity rate $\alpha_{i}$ [Fig.1(a)-(b)]. Contagion processes on these coupled structures networks are analytically studied. We sort the individuals into a series of classes, and obtain the threshold by analyzing the fixed point of the diffusion system. We show the epidemic thresholds on different coupling strategies, both for homogeneous and heterogeneous strong structure in the SADC model. Our analytical results in two coupling scenarios can recover and generalize the classical results in static and time-varying networks. These theoretical results are also verified by numerical simulations.

The following part is organized as follows. In section $\uppercase\expandafter{\romannumeral2}$, we introduce the SADC model. In section $\uppercase\expandafter{\romannumeral3}$, we analytically study the case where networks have homogeneous strong structure and time-varying weak structure.  In section $\uppercase\expandafter{\romannumeral4}$, we consider the case where networks have heterogeneous strong structure and time-varying weak structure. In section $\uppercase\expandafter{\romannumeral5}$, we point out that the average degree of weak structure has a threshold to effect the epidemic's outbreak, and explain why it may fail to stop contagion by destroying the weak structure.

\section{\label{sec2} The static and activity driven coupling model}

Firstly we show how to construct the weak and strong structures of a network in the static and activity driven coupling (SADC) model. We construct the weak structure by the activity driven model with initial activity probability $F(\alpha)$. The strong structure is generated according to a strong degree distribution $P(k_{1})$. In this way, each node has two kinds of degree, strong degree $k_{1}$ and weak degree $k_{2}$. In particular, in this model we allow the weak links coexist with strong links between nodes, this may correspond to the scenario that one may meet family member or close friends by chance in social environments.

\begin{figure}
\includegraphics[width=0.5\textwidth]{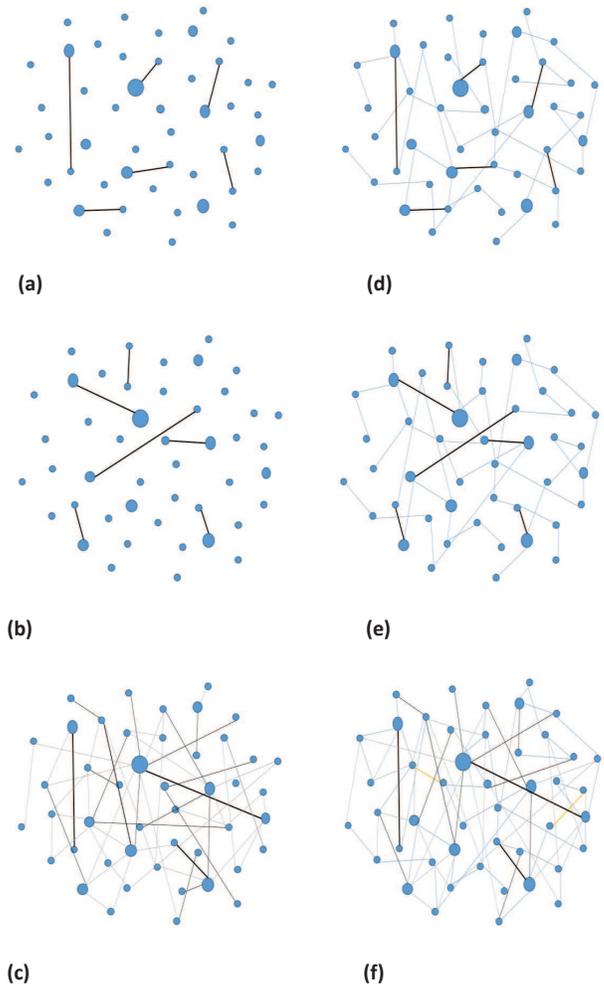}
\caption{(color online). Schematic representation of the static and activity driven (SADC) model. (a)-(b)Temporal weak structure at two different time $t_{1}$ and $t_{2}$,  which defined by activity driven model. (c)Integrated network consisted of weak links over a certain period of time. (d)-(e)Temporal SADC network at time $t_{1}$ and $t_{2}$, whose weak structure is showed in (a)-(b) and strong structure is same. (f) Integrated SADC network over a certain period of time. The size of each node describes its activity, while the width and color of each link describes the weight. Especially, the yellow one describes there has weak link and strong link simultaneously between two nodes.}\label{}
\end{figure}

The generation of SADC model can be illustrated as follows (Fig.1(d)-(e)): (i)Generating strong structure $G^{0}$: the strong structure is assigned with distribution $P(k_{1})$ and the strong links keep static. (ii)Generating networks $G^{t}$: At each discrete time step $t$, with probability $\alpha_{i}\Delta t$, each node $i$ on networks $G^{0}$ becomes active and generates $m$ weak links that connect with other nodes randomly. (iii) Contagion processes on networks $G^{t}$: the information spreads for one time step on networks $G^{t}$. (iv)At the next time step $t+\Delta t$, all the weak links in $G^{t}$ are deleted, and repeat the step 2 and step 3 above. In the following, without loss of generality, we set $\Delta t=1$.

We consider the SIS and SIR contagion\cite{Imm41} in our model. The basic $SIS$ ($SIR$) rules define a reaction of the type $S+I \rightarrow 2I$ with the probability $\beta$ per unit time and $I \rightarrow S$ ($I \rightarrow R$) with probability $\mu$ per unit time, which defines the contagion and recovery processes respectively. The epidemic threshold is a key concept of contagion progresses. It depends on $P(k_{1})$ in static networks\cite{Imm28,Imm40}, and is  decided by $F(\alpha)$ in activity driven networks\cite{Imm13}.

We adopt heavy-tailed distributions of activity, i.e $F(\alpha)\sim\alpha^{-\gamma}$, with activities restricted in the region $\alpha\in[\epsilon,1]$, where $\epsilon$ is a very small positive number, to avoid divergences for $\alpha \rightarrow 0$. Integrating the weak links in finite time $T$ (Fig.1(c)), its degree distribution follows $\frac{1}{Tm}F(\frac{k}{Tm}) $\cite{Imm13}, so the weak structure is heterogeneous. However, integrating SADC networks[Fig.1(f)], the degree distribution couples the heavy-tailed form with the function $P(k_{1})$. On the networks $G^{t}$, we define $\langle k_{1}\rangle$ and $\langle k_{2}\rangle$ as the average degree of the strong and the weak structure. So the average degree $\langle k\rangle$ equals to $\langle k_{1}\rangle$+$\langle k_{2}\rangle$. As the ratio of $\langle k_{2}\rangle/\langle k\rangle$ decreases from 1 to 0, the heterogeneous structure turns to homogeneous structure in SADC networks with homogeneous strong structures, while becomes the other heterogeneous structure in heterogeneous strong structures scenario. These characters in SADC networks differ from that in static networks and time-varying networks.

\section{\label{sec3}  contagion on SADC model with homogeneous strong structure}

At first, we consider the SIR model. The ratio of infected, susceptible and immune (removed) nodes with activity $\alpha$ at time $t$, among all of individuals, are denoted as $I^{t}_{\alpha}$, $S^{t}_{\alpha}$ and $R^{t}_{\alpha}$ respectively. We introduce the per contact infected probability as $\lambda$. In this sense, $\beta=\langle k\rangle\lambda$ gives the average contacts of per node with degree $\langle k\rangle$. For most of nodes, the degree of strong structure $k_{1} \approx \langle k_{1}\rangle$ in the homogeneous strong structure \cite{Imm40}. We write mean-field evolution of the ratio of infected individuals with activity $\alpha$ as
\begin{equation}
\begin{aligned}
{I}^{t+1}_{\alpha}  &= I^{t}_{\alpha}-\mu I^{t}_{\alpha} +\lambda m (P(\alpha)- I^{t}_{\alpha}-R^{t}_{\alpha})\alpha \int I^{t}_{\alpha'} d \alpha'\\
& +\lambda m (P(\alpha)-I^{t}_{\alpha}-R^{t}_{\alpha})\int\alpha'I^{t}_{\alpha'}d\alpha' \\
& +\lambda \langle k_{1}\rangle(P(\alpha)-I^{t}_{\alpha}-R^{t}_{\alpha})I^{t}
\end{aligned}
\end{equation}
where $P(\alpha)$ is the probability of individuals with activity rate $\alpha$. On the right side of Eq.(1), the second term represents the ratio of nodes which recovers from the class $I^{t}_{\alpha}$. The third term describes the ratio of infected individuals generated when nodes in the class $S^{t}_{\alpha}=P(\alpha)-I^{t}_{\alpha}-R^{t}_{\alpha}$ are active and connect with infected nodes via weak links. The forth term considers the ratio of infected individuals generated when nodes in the class $S^{t}_{\alpha}$ are connected by active infected nodes via weak links. Finally, the last term describes the ratio of infected individuals generated when nodes in the class $S^{t}_{\alpha}$ connect with infected nodes via strong links. In fact, the infected individuals are generally described as three parts: one is decided by strong structure, the other two are decided by weak structure.

Considering $R^{t}_{\alpha} \approx 0$ at the beginning of the spreading and ignoring the second order terms in $I^{t}$ ($I \thicksim 1/N$ at the beginning of epidemic spreading), we can rewrite $I^{t+1} = I^{t}-\mu I^{t}+\lambda m \langle\alpha\rangle I^{t}+\lambda m \theta^{t} + \lambda \langle k_{1}\rangle I^{t}$, where $I^{t} = \int I^{t}_{\alpha} d\alpha$ and $\theta^{t} = \int \alpha I^{t}_{\alpha} d\alpha$. By multiplying both sides of Eq.(1) with $\alpha$, we can obtain  $\theta^{t+1}  = \theta^{t} - \mu \theta^{t} + \lambda m \langle\alpha^{2}\rangle I^{t}+\lambda m \langle\alpha\rangle \theta^{t} + \lambda \langle\alpha\rangle \langle k_{1}\rangle I^{t}$. Considering the continuous progress, the master equations for $I^{t}$ and $\theta^{t}$ can be written as
\begin{eqnarray}
\partial_{t}I^{t} &=& -\mu I^{t} + \lambda m \langle\alpha\rangle I^{t} + \lambda m \theta^{t} + \lambda \langle k_{1}\rangle I^{t} \\
\partial_{t}\theta^{t} &=& -\mu \theta^{t} + \lambda m \langle\alpha^{2}\rangle I^{t} + \lambda m \langle\alpha\rangle \theta^{t} + \lambda \langle\alpha\rangle \langle k_{1}\rangle I^{t}.
\end{eqnarray}
The above equations for $I^{t}$ and $\theta^{t}$ show an epidemic outbreak if and only if the dominant eigenvalue of the corresponding matrix is larger than $1$. Actually, the epidemic threshold is given by
 \begin{eqnarray}\nonumber
   \frac{\lambda}{\mu} &\geq& \varphi^{Ho} \\
   &\equiv& \frac{1}{m \langle\alpha\rangle+\frac{1}{2}\langle k_{1}\rangle+\sqrt{\frac{1}{4}\langle k_{1}\rangle^{2}+m^{2}\langle\alpha^{2}\rangle+m\langle k_{1}\rangle\langle\alpha\rangle}}.
 \end{eqnarray}
 For SIS model, $R^{t}_{\alpha}=0$ in Eq.(1), the threshold also can be given by Eq.(4). The average weak degree of node $i$  with activity rate $\alpha_{i}$ is $\langle k_{2,i}\rangle=m\alpha_{i}+m \langle\alpha\rangle$, thus, the average weak degree $\langle k_{2}\rangle=2m\langle\alpha\rangle$. In this way, the epidemic threshold can be also described as $\frac{\beta}{\mu} \geq \langle k\rangle\varphi^{Ho}$, where $\langle k\rangle=2m\langle \alpha\rangle+\langle k_{1}\rangle$. Suppose there's no weak links in the SADC model (i.e.$m \equiv 0$), Eq.(4) recovers the classical result $\frac{\lambda}{\mu}\geq\varphi^{Ho}=1/\langle k_{1}\rangle$\cite{Imm39}. Meanwhile, when $\langle k_{1}\rangle\equiv0$, which means there's no strong structure in SADC model, it also recovers the classical result in time-vary networks $\frac{\beta}{\mu} \geq \langle k\rangle\varphi^{Ho}\equiv2\langle \alpha\rangle/(\langle \alpha\rangle+\sqrt{\langle \alpha^{2}\rangle})$ \cite{Imm13}.
 \begin{figure}
\includegraphics[width=0.5\textwidth]{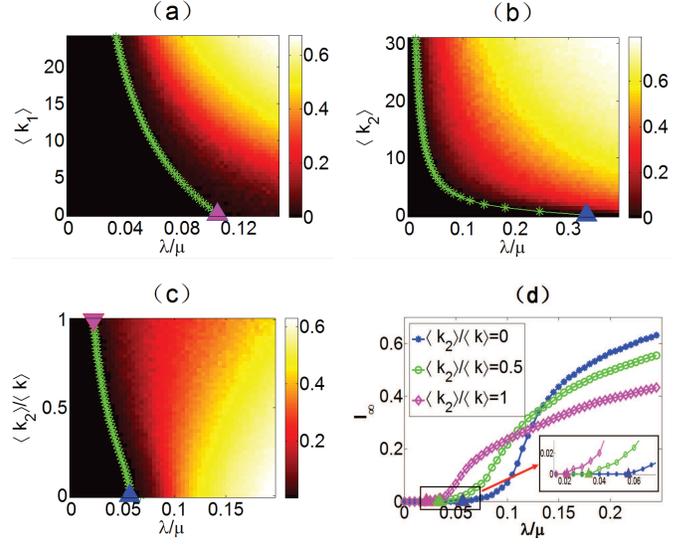}
\caption{(color online). A SIS process on SADC networks with homogeneous strong structure. (a)The phase diagram of $I_{\infty}$ for each pair of $\langle k_{1}\rangle$ and $\lambda/\mu$ with $\langle k_{2}\rangle=2$. (b)The phase diagram of $I_{\infty}$ for each pair of $\langle k_{2}\rangle$ and $\lambda/\mu$ with $\langle k_{1}\rangle=2$. (c)The phase diagram of $\langle k_{2}\rangle/\langle k\rangle$ and $\lambda/\mu$ with $\langle k\rangle=17.62$. The green lines in the three phase spaces represent the critical value in our model. (d)Comparison of the stationary state$(I_{\infty})$ for a SIS model with $\langle k\rangle=17.62$ but different $\langle k_{2}\rangle/\langle k\rangle$ on SADC networks. The triangle represents the critical value defined by our model. The purple triangles recover the classical conclusion in the activity driven model and the blue ones recover the classical prediction in static networks. Considering $N=10^{4}$, $\epsilon=10^{-3}$, activity distributed as $F(\alpha)=\alpha^{-2.2}$, strong structure is ER model, each plot is the average of $10^{2}$ independent realizations started with $1\%$ of random infected seeds.}\label{}
\end{figure}

Based on this framework, we study the SIS process on SADC model with three coupling strategies. In the first case, the average weak degree $\langle k_{2}\rangle$ is fixed. We then simulate the contagion process and contour plot the average asymptotic density of infected nodes $I_{\infty}$ in $10^{2}$ independent realizations as a function of both $\langle k_{1}\rangle$ and $\lambda/\mu$ (as shown in Fig.2(a)). We show that the phase diagram of the diffusion process is divided into two different regions by a green line which represents $\varphi^{Ho}$ as derived by Eq.(4). In the second case, by keeping the strong structure $\langle k_{1}\rangle$ as a constant, we also contour plot $I_{\infty}$ as a function of   $\langle k_{2}\rangle$ and $\lambda/\mu$ in Fig.2(b), and the green line separates the phase diagram into two parts clearly. In the third case, in Fig.2(c), we consider the average degree $\langle k\rangle$ as a constant, while the ratio $\langle k_{2}\rangle/\langle k\rangle$ is varying. We contour plot $I_{\infty}$  as a function of $\langle k_{2}\rangle/\langle k\rangle$ and $\lambda/\mu$. The same as the two above figures, the green line separates the exploding state from un-exploding state in the phase diagram.

Another important point shown in Fig.2(c) is that a larger value of $\langle k_{2}\rangle/\langle k\rangle$ may result in a smaller epidemic threshold but a much slower diffusion speed. To show this more clearly, we choose three different values of $\langle k_{2}\rangle/\langle k\rangle$ and plot the asymptotic density of infected individuals $I_{\infty}$ as a function of $\lambda/\mu$ in Fig.2(d). We show that the weak structure can make the epidemics outbreak in advance, but slows the spreading speed after epidemic outbreak. In fact, when $\langle k_{2}\rangle/\langle k\rangle$ getting larger, there will be more weak links in our SADC networks. Thus, the contagion process spreads mainly via weak structure. In the weak structure, nodes with high activity connect the infected nodes more easily, which makes them much more easily to become infected nodes. In this way, the infected nodes with high activity easily become active and accelerate the diffusion, which makes the epidemic threshold much smaller. On the other hand, since there is a large number of nodes with low activity in the weak structure, it is not easy for them to infect other nodes before recovery. Thus, the contagion process will be slowed after the outbreak when $\langle k_{2}\rangle/\langle k\rangle$ increases.

Meantime, our result also recovers the classical results of the two limit cases of SADC networks exactly, which is shown by purple triangles and blue triangles in Fig.2. For the
SIR model, our framework can also predict the epidemic threshold exactly(Fig.3(a)).

\begin{figure}
\includegraphics[width=0.5\textwidth]{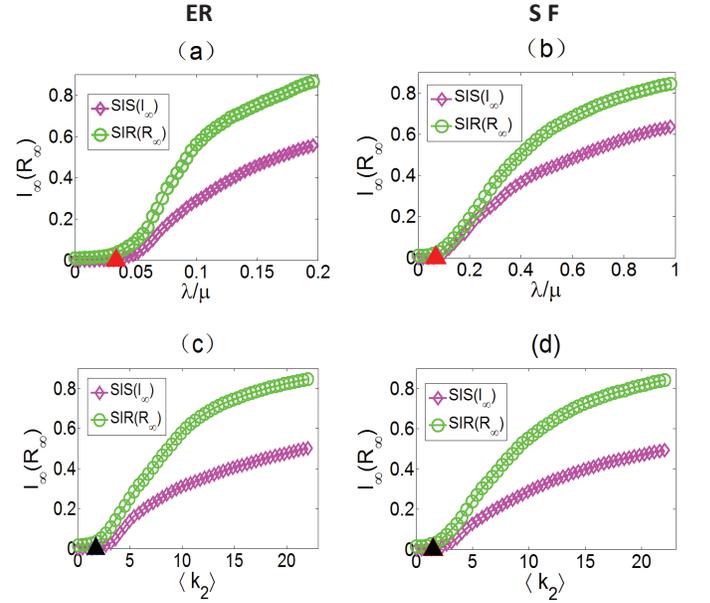}
\caption{(color online). The asymptotic density of a SIR$(R_{\infty})$ and a SIS$(I_{\infty})$, as a function of $\lambda/\mu$ or $\langle\emph{k}_{2}\rangle$ respectively. (a)(c)The strong structure is ER model. (b)(d) The strong structure is scale-free. The green circles represent the median of $R_{\infty}$ and purple rhombus show the median of $I_{\infty}$. The red triangle represents the analytical prediction of the threshold according to Eq.4 and Eq.11, the black triangle represents the analytical prediction of threshold defined by the Eq.12 and Eq.13. (a)$\langle \emph{k}_{1}\rangle=8.62$, $\langle \emph{k}_{2}\rangle=8.99$. (b)$\langle \emph{k}_{1}\rangle=2.79$, $\langle \emph{k}_{2}\rangle=1.80$, $F(k_{1})\sim k_{1}^{-2.5}$. (c)$\langle \emph{k}_{1}\rangle=3$, $\lambda/\mu=0.15$. (d)$\langle \emph{k}_{1}\rangle=2$, $F(k_{1})\sim k_{1}^{-2.99}$, $\lambda/\mu=0.15$. Considering $10^{4}$ nodes, $\epsilon=10^{-3}$, and a power law distribution of activity $F(\alpha)\sim\alpha^{-2.2}$, the plot is the average of $10^{2}$ independent realizations and each one of them starts with $1\%$ of random infected seeds.}\label{}
\end{figure}

\section{\label{sec4}  contagion on the SADC model with heterogeneous strong structure}

In the following part, we consider the strong structure to be heterogeneous in SADC model. In this situation, we suppose $k_{1}$ differs a lot for each node. Let us denote $I^{t}_{\alpha,k_{1}}$, $S^{t}_{\alpha,k_{1}}$ and $R^{t}_{\alpha,k_{1}}$ as the ratio of infected, susceptible and immune (removed) nodes, respectively, with activity $\alpha$ and strong degree $k_{1}$ at time $t$, among all individuals. Then we can describe the ratio of infected individuals $I^{t}_{\alpha,k_{1}}$ as
\begin{eqnarray} \nonumber
I^{t+1}_{\alpha,k_{1}}  &=&  I^{t}_{\alpha,k_{1}} - \mu I^{t}_{\alpha,k_{1}} \\ \nonumber
  && + \lambda m (P(\alpha,k_{1}) - I^{t}_{\alpha,k_{1}} - R^{t}_{\alpha,k_{1}}) \alpha \int\int I^{t}_{\acute{\alpha},k_{1}} d\acute{\alpha}dk_{1} \\\nonumber
   && +\lambda m (P(\alpha,k_{1}) - I^{t}_{\alpha,k_{1}} - R^{t}_{\alpha,k_{1}}) \int\int\acute{\alpha} I^{t}_{\acute{\alpha},k_{1}} d\acute{\alpha}dk_{1} \\
  && +\lambda k_{1} (P(\alpha,k_{1})-I^{t}_{\alpha,k_{1}} - R^{t}_{\alpha,k_{1}}) \Psi_{k_{1}}^{t}
\end{eqnarray}
where $P(\alpha,k_{1})$ is the probability of a random selected individual of activity rate $\alpha$ and strong links $k_{1}$. $\Psi_{k_{1}}^{t}$ is the the probability that a node with strong links $k_{1}$ connects the infected individuals via a strong link. In fact, $\Psi_{k_{1}}^{t}$ equals $I^{t}$ in homogeneous coupling scenario. On the right side of Eq.(5), the second term is the ratio of nodes that recover from the class of $I^{t}_{\alpha,k_{1}}$. The third term describes the ratio of infected individuals generated when nodes in the class $S^{t}_{\alpha,k_{1}}=P(\alpha,k_{1})-I^{t}_{\alpha,k_{1}}-R^{t}_{\alpha,k_{1}}$ are active and connect infected ones via weak links. The fourth term considers the ratio of infected individuals generated when nodes in the class $S^{t}_{\alpha,k_{1}}$ connected by active infected ones via weak links. The last term describes the ratio of infected individuals generated when nodes in the class $S^{t}_{\alpha,k_{1}}$ connect infected ones in strong structure.

Here we consider the SIS model first, i.e. $R^{t}_{\alpha,k_{1}}=0$, and suppose weak structure and strong structure are independent, then $P(\alpha,k_{1})=P(\alpha)P(k_{1})$. We check the fixed point of the diffusion system (see details in appendix) and get
\begin{equation}\label{6}
  I^{*}_{k_{1}} = \frac{\lambda m P(k_{1}) \langle\alpha\rangle I^{*} + \lambda m P(k_{1}) \theta^{*} + \lambda k_{1} P(k_{1}) \Psi^{*}_{k_{1}}}{\mu + \lambda k_{1}\Psi^{*}}
\end{equation}
Where $I^{*}=\int I^{*}_{k_{1}} dk_{1}$ and $\theta^{*}=\int\int\alpha I^{*}_{\alpha,k_{1}} d\alpha dk_{1}$. $\Psi^{*}_{k_{1}}$ represents the critical connecting probability $\Psi_{k_{1}}^{t}$ when the system reaches the fixed point.
We suppose that the strong structure has no degree correlations, i.e., links between any random selected node pairs are irrelevant with their degree, so the values of $\Psi^{*}_{k_{1}}$ are all supposed to be equal to $\Psi^{*}$. Meantime, the probability $P(s|k_{1})$, describing the node with strong links $k_{1}$ connects the node with strong links $s$, equals $sP(s)/\langle k_{1}\rangle$. So $\Psi^{*}$ must satisfy
\begin{eqnarray} \nonumber
  \Psi^{*} &=& \int P(s|k_{1}) \frac{I^{*}_{s}}{P(s)} = \frac{1}{\langle k_{1}\rangle} \int s I^{*}_{s}\\
    &=& \frac{1}{\langle k_{1}\rangle} \int s P(s)(\frac{\lambda s\Psi^{*} + \lambda m \langle\alpha\rangle I^{*} + \lambda m \theta^{*}}{\mu + \lambda s \Psi^{*}}).
\end{eqnarray}

According to the Eq.(5), when the system reaches the fixed point, $I^{*} $ and $\theta^{*}$ (see details in appendix) can be written as
\begin{eqnarray}
 I^{*} &=& \frac{\lambda \mu \langle k_{1}\rangle \Psi^{*}}{(\mu - \lambda m \langle\alpha\rangle)^{2} - \lambda^{2} m^{2} \langle\alpha^{2}\rangle}\\
\theta^{*} &=& \frac{(\lambda \mu \langle k_{1}\rangle \langle\alpha\rangle + \lambda^{2} m \langle k_{1}\rangle (\langle\alpha^{2}\rangle - \langle\alpha\rangle^{2})) \Psi^{*}} {(\mu - \lambda m \langle\alpha\rangle)^{2} - \lambda^{2} m^{2} \langle\alpha^{2}\rangle}.
\end{eqnarray}

In fact, if we substitute Eq.(7) into the self-consistent equation  $\Psi^{*} = H(\Psi^{*})$, it can be shown that $\partial^{2}_{\Psi}H(\Psi) \leq 0$ and $\Psi^{*} = 0$ is a solution. If epidemic outbreaks, a positive value $\Psi^{*}$ must satisfy $\Psi^{*} = H(\Psi^{*})$. Thus, if the epidemic outbreaks, $\partial_{\Psi}H(0)\geq1$, then
\begin{equation}\label{10}
\frac{1}{\langle k_{1}\rangle}\frac{\lambda}{\mu} \int sP(s)(s + (m \langle\alpha\rangle \partial_{\Psi^{*}}I^{*} + m \partial_{\Psi^{*}}\theta^{*})\mid_{\Psi^{*}=0}) - 1 \geq 0
\end{equation}
where $\partial_{\Psi^{*}}I^{*}\mid_{\Psi^{*}=0}$ and $\partial_{\Psi^{*}}\theta^{*}\mid_{\Psi^{*}=0}$ are given by Eq.(8) and Eq.(9) respectively (see details in appendix). In fact, without the weak structure, Eq.(10) recovers the classical conclusion $\lambda/\mu\geq\langle k_{1}\rangle/\langle k_{1}^{2}\rangle$ \cite{Imm39}. Without the strong structure, this equation can also recover the same result for the time-varying networks \cite{Imm13}.

To simplify, we denote the left-hand side of Eq.(10) as $\Lambda(\lambda/\mu)$. The epidemic will outbreak when $\Lambda(\lambda/\mu)>0$. In fact, $\frac{\langle k_{1}^2\rangle}{\langle k_{1}\rangle} \geq 1$, when $\frac{\lambda}{\mu} \geq 1$, then the epidemic will outbreak according to Eq.(10). Next we consider the case $\frac{\lambda}{\mu} \in (0,1)$. By making the Laurent expansion $\Lambda(\frac{\lambda}{\mu}) = \sum^{5}_{0}f_{i}(\frac{\lambda}{\mu})^{i}+o((\frac{\lambda}{\mu})^{5})$, we denote $\widetilde{\Lambda}(\frac{\lambda}{\mu}) = \sum^{5}_{0}f_{i}(\frac{\lambda}{\mu})^{i}$ (see details in appendix). Considering $\langle k_{2}\rangle > 0$ and $\langle k_{1}\rangle>0$ in SADC networks, it can be proved that $f_{5}>0, \widetilde{\Lambda}(0)<0$ and $ \widetilde{\Lambda}'(\frac{\lambda}{\mu})>0$ as $\frac{\lambda}{\mu}>0$, so the equation $\widetilde{\Lambda}(\frac{\lambda}{\mu})=0$ has one and only one positive solution. Thus, the epidemic threshold can be written as
\begin{equation}\label{11}
  \frac{\lambda}{\mu} \geq \varphi^{He} \approx \phi_{max}
\end{equation}
where $\varphi^{He}$ represents the epidemic threshold in SADC model with heterogeneous strong structure, $\phi_{max}$ is the positive root of $\widetilde{\Lambda}(\frac{\lambda}{\mu})=0$. As $\frac{\langle k_{1}^{2}\rangle}{\langle k_{1}\rangle}\rightarrow\infty$, the value $\phi_{max} \rightarrow 0$ (i.e. $\varphi^{He} \rightarrow 0$). When the SADC networks with heterogeneous strong structure is large enough, the epidemic threshold is close to zero, which is in accordance with the classic result of static heterogeneous networks.
 \begin{figure}
\includegraphics[width=0.5\textwidth]{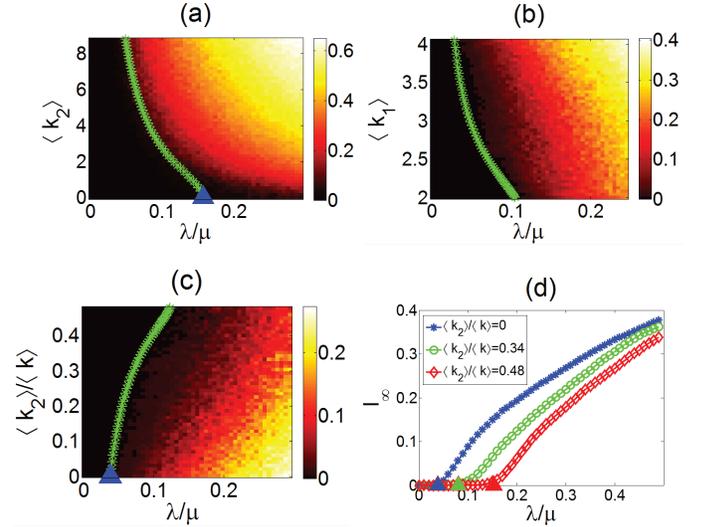}
\caption{(color online). A SIS process on SADC network with heterogeneous strong structure. (a)The phase diagram of $I_{\infty}$ for each pair of $\langle k_{2}\rangle$ and $\lambda/\mu$ with $\langle k_{1}\rangle=2.1$. (b)The phase diagram of $I_{\infty}$ for each pair of $\langle k_{1}\rangle$ and $\lambda/\mu$ with $\langle k_{2}\rangle=1.8$. (c)The phase diagram of $I_{\infty}$ for each pair of $\langle k_{2}\rangle/\langle k\rangle$ and $\lambda/\mu$ with $\langle k\rangle=3.93$. The green lines in the three phase diagrams represent the critical value in our model. (d)Comparison of the stationary state $(I_{\infty})$ for a SIS model with $\langle k\rangle=3.92$ but different $\langle k_{2}\rangle/\langle k\rangle$ on SADC networks. The triangle represents the critical value defined by our model. The blue triangles recover the classical prediction in static networks. Considering $N=10^{4}$, $\epsilon=10^{-3}$, activity distributed as $F(\alpha)=\alpha^{-2.2}$, strong structure is SF model, each plot is the average of $10^{2}$ independent realizations started with $1\%$ of random infected seeds.}\label{}
\end{figure}

To verify, we simulate the SIS process on SADC networks where the strong structure is of scale-free type. We show that our theoretical prediction gives the epidemic thresholds correctly and recovers all the classical conclusions. In Fig.4(a), with the strong structure fixed, we contour plot the average asymptotic density of infected nodes $I_{\infty}$ as a function of $\langle k_{2}\rangle$ and $\lambda/\mu$. The phase diagram of $\emph{I}_{\infty}$ can be identified as two distinct regions by a green line as derived by Eq.(11). We then fix $\langle k_{2}\rangle$ and change the strong structure in Fig.4(b), we can see the phase diagram can be also separated into two regions, i.e. exploding and un-exploding state, by the green line representing the theoretical prediction threshold. In Fig.4(c), we consider the average degree $\langle k\rangle$ as a constant while the ratio $\langle k_{2}\rangle/\langle k\rangle$ increases, again, the green line obtained from our theoretical framework correctly separates the phase diagram into two regions. Different from the homogeneous coupling (Fig.2(c)-(d)), as the ratio $\langle k_{2}\rangle/\langle k\rangle$ increases, the contagion threshold becomes larger and larger, which is shown more clearly in Fig.4(d). In fact, nodes with large degrees are much more easily to be infected in SF networks, which accelerates the contagion process. However, as the ratio $\langle k_{2}\rangle/\langle k\rangle$ increases, the contagion spreads mainly via weak structure, the accelerating effect weakens. Moreover, infected nodes with low activity can hardly infect other nodes before recovery, which makes the epidemic spreading slowly in weak structure, and leads epidemics outbreak later.

Our model also can be used to study the SIR model. Before the epidemics outbreak, the ratio of recovery individuals $R^{t}_{\alpha,k_{1}}$ approaches $0$ in Eq.(5). Thus, the analytical framework about SIR is same with SIS model. In Fig.3(b), we simulate the SIS and SIR process respectively. The two processes have similar epidemic threshold predicted by the Eq.(11).

\section{\label{sec5}  controlling contagion by destroying weak structure}

Based on our model, in this section, we study how the weak structure influence the epidemics on SADC networks. In fact, the average weak degree has a threshold in controlling the contagions, according to the Eq.(4) and Eq.(11). It can be written as
\begin{eqnarray}
   \langle k^{c}_{2}\rangle &=& \frac{2 \mu - 2 \sqrt{\mu^{2} \frac{\langle\alpha^{2}\rangle}{\langle\alpha\rangle^{2}} + \lambda \mu \langle k_{1}\rangle (1 - \frac{\langle\alpha^{2}\rangle}{\langle\alpha\rangle^{2}})}}{\lambda (1 - \frac{\langle\alpha^{2}\rangle}{\langle\alpha\rangle^{2}})}  \\
  \langle k^{c}_{2}\rangle &=& 2 \langle\alpha\rangle m^{c}
\end{eqnarray}
where Eq.(12) and Eq.(13) gives the homogeneous and heterogeneous coupling scenario, respectively. We consider $\widetilde{\Lambda}(m)=\widetilde{\Lambda}(\frac{\lambda}{\mu})$, and $m^{c}$ is the largest real root of $\widetilde{\Lambda}(m)=0$. If $\langle k_{2}^{c}\rangle>0$ and $\langle k_{2}\rangle<\langle k_{2}^{c}\rangle$, the contagion will not breakout, which indicates that the contagion will never outbreak no matter we destroy the weak structure or not. However, if the value $\langle k_{2}^{c}\rangle$ is not real or $\langle k_{2}^{c}\rangle<0$, the epidemics will outbreak, even we destroy the weak structure completely.

In Fig.5, We contour plot the average asymptotic density of infected nodes $\emph{I}_{\infty}$ as a function of $\langle k_{2}^{c}\rangle$ and $\langle k_{1}\rangle$. The phase diagram of the diffusion process is divided in two regions by the green dotted line which represents $\langle k_{2}^{c}\rangle$ as derived by Eq.(12) or Eq.(13). By comparing Fig.5(a) and Fig.5(b), as the value $\lambda/\mu$ increases, the threshold $\langle k^{c}_{2}\rangle$ decreases. Compared with homogeneous coupling, the threshold $\langle k^{c}_{2}\rangle$ is smaller and the epidemics outbreak in advance on heterogeneous coupling scenario(Fig.5(b)(c)).

Furthermore, we point out that destroying the weak structure while the epidemic outbreaks can delay the spreading effectively, however, this strategy may not always make the epidemics die out. If $\langle k^{c}_{2}\rangle>0$ (in Fig5(d)), destroy the weak structure is effective to slow down the epidemic spreading and make the epidemic die out eventually. However, when $\langle k^{c}_{2}\rangle<0$ (in Fig5(e)(f)), destroy weak structure can not stop the spreading. We also simulate the SIR process in Fig.3(c)(d). We show the value of $\langle k_{2}^{c}\rangle$ plays an important role in controlling the epidemics when $\langle k_{2}^{c}\rangle>0$.
 \begin{figure}
\includegraphics[width=0.5\textwidth]{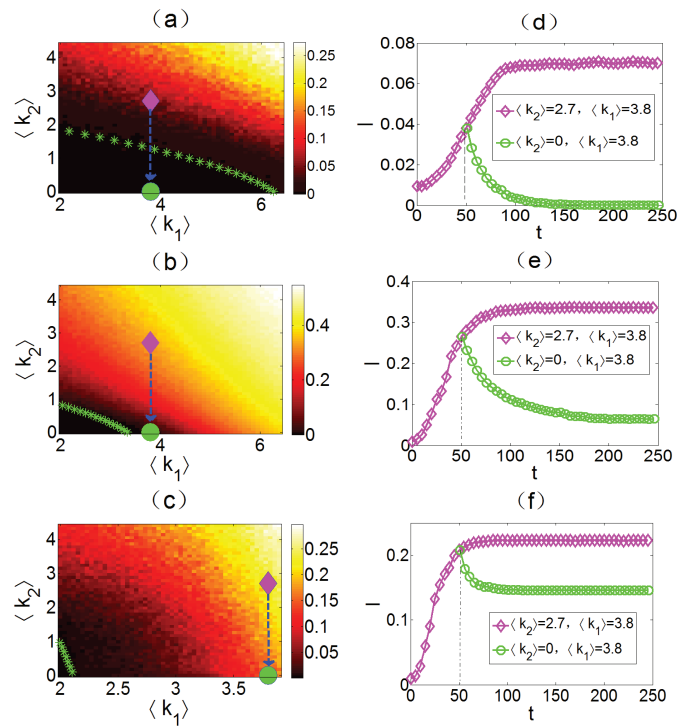}
\caption{(color online). (a)-(c)Phase diagram of the asymptotic density$(I_{\infty})$ for a SIS process on SADC network. The green dot line in phase diagrams represents the $\langle k_{2}^{c}\rangle$ defined by our model. (d)-(f) Comparison of controlling efficiency for a SIS model by destroying the weak structure, where $\langle k_{1}\rangle=3.82$, red purple rhombus represents $\langle k_{2}\rangle=2.7$ and the green circle represents $\langle k_{2}\rangle=0$. (a)(d)$\lambda/\mu=0.16$, the strong structure is ER model. (b)(e)$\lambda/\mu=0.3$, the strong structure is ER model. (c)(f)$\lambda/\mu=0.16$, the strong structure is scale-free. Considering $\emph{N}=10^{4}$, $\epsilon=10^{-3}$, activity distributed as $F(\alpha)=\alpha^{-2.2}$, each plot is the average of $10^{2}$ independent realizations started with $1\%$ of random infected seeds.}\label{}
\end{figure}

\section{\label{sec6}  conclusions}

In summary, we have studied these temporal networks that involving two structures with different time scales, strong structure and weak structure. We propose the static and activity driven coupling (SADC) model to describe the coupling between two structures. Within this framework, we thoroughly study the contagion processes on the static and time-varying coupling networks. Our analytical framework has gotten the epidemic thresholds of SIS and SIR model on SADC both analytically and numerically with various coupling strategies, both for homogeneous and heterogeneous strong structure. Theoretical thresholds obtained from SADC model can both recover and generalize the classical results in static and time-varying networks.

In particular, a huge difference is observed On SADC networks between homogeneous coupling and heterogeneous coupling. On homogeneous coupling scenario, keeping same average degree, bigger value $\langle k_{2}\rangle/\langle k\rangle$ can make the epidemics outbreak much more easily, while the spreading speed is slightly slower after epidemic outbreaks. Nevertheless, this phenomenon doesn't happen on heterogeneous coupling scenario, where the threshold delays. We conclude that weak structure shows different functions in two scenarios. On the one hand, the very few nodes with large activity accelerate the contagion. On the other hand, large number of nodes with low activity decelerate the spreading. The accelerated function is the main effect in homogeneous coupling, while the decelerated one is the main effect in heterogeneous coupling. Furthermore, our results promote our understanding of why most common epidemics can break out in reality, even when we control weak structure in social networks. Generally, if $\langle k_{2}^{c}\rangle>0$, the prevention by destroying the weak structure is effective. Otherwise, the action will fail to stop the contagion.

\section{\label{sec7} ACKNOWLEDGMENTS}
We thank the referees for helpful comments. This work is partially supported by the NSFC No. 11201017 and 11290141, Cultivation Project of NSFC (No. 91130019).

\section{\label{sec8}  Appendix: further explanations of section $\uppercase\expandafter{\romannumeral4}$}
In this paper, we suppose weak structure and strong structure are independent, and consider the strong structure has no degree correlations, then we note $\Psi^{t}_{k_{1}}=\Psi^{t}$. By ignoring the second order terms in $I^{t}$ and integrating both sides of Eq.(5) by $\alpha$, we obtain $I_{k_{1}}^{t+1} = I_{k_{1}}^{t} - \mu I_{k_{1}}^{t} + \lambda m P(k_{1}) \langle\alpha\rangle I^{t} + \lambda m P(k_{1}) \theta^{t} + \lambda k_{1}(P(k_{1}) - I_{k_{1}}^{t})\Psi^{t}$, where $I_{k_{1}}^{t} = \int I_{\alpha,k_{1}}^{t} d\alpha$,
$I^{t} = \int\int I_{\alpha,k_{1}}^{t} d\alpha dk_{1}$ and $\theta^{t} = \int\int\alpha I_{\alpha,k_{1}}^{t}d\alpha dk_{1}$. By integrating both sides of Eq.(5) by $\alpha$ and $k_{1}$ and ignoring the second order terms in $I^{t}$, we can get $I^{t+1}=I^{t}-\mu I^{t}+\lambda m\langle\alpha\rangle I^{t}+\lambda m\theta^{t}+\lambda\langle k_{1}\rangle\Psi^{t}$. By multiplying both sides of Eq.(5) by $\alpha$ and integrating by $\alpha$ and $k_{1}$, and ignoring the second order terms in $I^{t}$, we obtain $\theta^{t+1} = \theta^{t} -\mu\theta^{t}+\lambda m\langle\alpha^{2}\rangle I^{t}+\lambda m\langle\alpha\rangle\theta^{t}+\lambda\langle k_{1}\rangle\langle\alpha\rangle\Psi^{t}$. Considering the continuous progress, we obtain
\begin{eqnarray*}
  \partial_{t}I_{k_{1}}^{t} &=& -\mu I_{k_{1}}^{t} + \lambda m P(k_{1}) \langle\alpha\rangle I^{t} \\
   &&+ \lambda m P(k_{1}) \theta^{t} + \lambda k_{1}(P(k_{1} - I_{k_{1}}^{t})\Psi^{t} \\
 \partial_{t} I^{t} &=& -\mu I^{t} + \lambda m \langle\alpha\rangle I^{t} + \lambda m \theta^{t} + \lambda \langle k_{1}\rangle \Psi^{t}\\
  \partial_{t}\theta^{t} &=& -\mu \theta^{t} + \lambda m \langle\alpha^{2}\rangle I^{t} + \lambda m \langle\alpha\rangle \theta^{t} + \lambda\langle k_{1}\rangle \langle\alpha\rangle \Psi^{t}.
\end{eqnarray*}

Considering the fix point of this system, we can get Eq.(6), Eq.(8) and  Eq.(9). The value of $\Psi^{*}$ should meet Eq.(7), namely $\Psi^{*}=H(\Psi^{*})$, where
\begin{equation*}
  H(\Psi) = \frac{1}{\langle k_{1}\rangle} \int s P(s)(\frac{\lambda s \Psi + \lambda m \langle\alpha\rangle I^{*} + \lambda m \theta^{*}}{\mu + \lambda s \Psi}).
\end{equation*}
According to Eq.(8) and Eq.(9), we can get
\begin{eqnarray*}
  \partial_{\Psi^{*}=0}I^{*} &=& \frac{\lambda \mu \langle k_{1}\rangle}{(\mu - \lambda m \langle\alpha\rangle)^{2} - \lambda^{2} m^{2} \langle\alpha^{2}\rangle} \\
  \partial_{\Psi^{*}=0}\theta^{*} &=& \frac{\lambda^{2} m \langle k_{1}\rangle(\langle\alpha^{2}\rangle - \langle\alpha\rangle^{2}) + \lambda \mu \langle k_{1}\rangle \langle\alpha\rangle}{(\mu - \lambda m \langle\alpha\rangle)^{2} - \lambda^{2} m^{2} \langle\alpha^{2}\rangle}
\end{eqnarray*}
It can be proved that $\partial^{2}_{\Psi}H(\Psi)\leq0$ and $0=H(0)$. If a positive value $\Psi^{*}$ meets the function, it must be verified that $\partial_{\Psi}H(0)\geq1$, namely Eq.(10). Meantime,
\begin{equation*}
  \frac{\langle k_{1}\rangle}{\langle k^{2}_{1}\rangle + \int s P(s)((m \langle\alpha\rangle \partial_{\Psi^{*}}I^{*} + m \partial_{\Psi^{*}}\theta^{*})\mid_{\Psi^{*}=0})} < 1.
\end{equation*}
If $\lambda\geq\mu$, then the contagion outbreaks based on the Eq.(10), so we consider the value $\frac{\lambda}{\mu}\in(0,1)$. Eq.(10) is a implicit function of $\frac{\lambda}{\mu}$, we analyse its laurent expansion. We let $\Lambda(\lambda/\mu)=\partial_{\Psi}H(0)-1$, then
$\Lambda(\frac{\lambda}{\mu})=\sum^{5}_{0}f_{i}(\frac{\lambda}{\mu})^{i}+o((\frac{\lambda}{\mu})^{5})$, where $f_{0}=-1, f_{1}=\frac{\langle k_{1}^{2}\rangle}{\langle k_{1}\rangle}, f_{2}=2m\langle\alpha\rangle\langle k_{1}\rangle, f_{3}=m^{2}\langle k_{1}\rangle\langle\alpha^{2}\rangle+3m^{2}\langle k_{1}\rangle\langle\alpha\rangle^{2}, f_{4}=8m^{3}\langle\alpha\rangle\langle k_{1}\rangle\langle\alpha^{2}\rangle+8m^{3}\langle k_{1}\rangle\langle\alpha\rangle^{3}, f_{5}=6m^{4}\langle k_{1}\rangle\langle\alpha^{2}\rangle^{2}+60m^{4}\langle k_{1}\rangle\langle\alpha^{2}\rangle\langle\alpha\rangle^{2}+30m^{4}\langle\alpha\rangle^{4}\langle k_{1}\rangle$. We consider $\widetilde{\Lambda}(\frac{\lambda}{\mu})=\sum^{5}_{0}f_{i}(\frac{\lambda}{\mu})^{i}$. Considering $\langle k_{2}\rangle>0$ and $\langle k_{1}\rangle>0$ in SADC networks, $f_{5}>0, \widetilde{\Lambda}(0)<0$ and $ \widetilde{\Lambda}'(\frac{\lambda}{\mu})>0$ as $\frac{\lambda}{\mu}>0$, the function $\widetilde{\Lambda}(\frac{\lambda}{\mu})=0$ has one and only one positive root, noted as $\phi_{max}$. So the epidemic threshold can be written as
\begin{eqnarray*}
  \frac{\lambda}{\mu} \geq \varphi^{He} \approx \phi_{max}
\end{eqnarray*}
Specially, if $\langle k_{2}\rangle=0$, we can get $\phi_{max}=\langle k_{1}\rangle/\langle k_{1}^2\rangle$, which is the classical result in static networks.

\bibliography{apssamp}

\end{document}